\documentclass[aps,prb,showpacs,twocolumn]{revtex4}
\usepackage{amsfonts}
\usepackage{amssymb}
\usepackage{amsmath}
\usepackage{graphicx}
\usepackage{dcolumn}
\usepackage{bm}

\begin{document}

\title{First principles study on the spin transfer torques}
\author{Shuai Wang}
\affiliation{State Key Laboratory for Surface Physics, Institute of Physics, Chinese
Academy of Sciences, P.O. Box 603, Beijing 100080, China}
\author{Yuan Xu}
\affiliation{State Key Laboratory for Surface Physics, Institute of Physics, Chinese
Academy of Sciences, P.O. Box 603, Beijing 100080, China}
\author{Ke Xia}
\affiliation{State Key Laboratory for Surface Physics, Institute of Physics, Chinese
Academy of Sciences, P.O. Box 603, Beijing 100080, China}
\date{\today }

\begin{abstract}
An efficient first principles method was developed to calculate spin
transfer torques in layered system with noncollinear magnetization. The
complete scattering wave function is determined by matching the wave
function in the scattering region with the Bloch states in the leads. The
spin transfer torques are obtained with aid of the scattering wave function.
We applied our method to the ferromagnetic spin valve and found that the
material (Co, Ni and Ni$_{80}$Fe$_{20}$) dependence of the spin transfer
torques could be well understood by the Fermi surface. Ni has much longer
spin injection penetration length than Co. Interfacial disorder is also
considered. It is found that the spin transfer torques could be enhanced by
the interfacial disorder in some system.
\end{abstract}

\maketitle

\section{Introduction}

Spin angular momentum can be transferred by the flowing electrons from one
ferromagnetic (FM) material to another FM material, which is so-called spin
transfer torques (STT) introduced by Slonczewski\cite{J.Slonc96} and Berger%
\cite{Berger96}. Those two seminal studies have shown that the dynamics of
magnetization in FM material could be dominated by the spin torques carried
by electric current. The excitation of coherent precession of magnetization
and spin wave were predicted. The STT was soon identified in the experiments%
\cite{experiment} by clear observation of the magnetization switching in FM
spin valve, which excites great interests in experiment and theory\cite%
{sun00,zhangsc98,Waintal,Brataas_circuit,MDstiles02,stiles02,Edwards05,PMHaney}%
.

The theories\cite{Waintal,Brataas_circuit,MDstiles02,stiles02} combining the
quantum treatment of the interface scattering and the Boltzmann-like
treatment of the bulk scattering work reasonable well with the experiments
of metallic system. However, recent experiments on the tunnelling system\cite%
{Fuchs} and magnetic domain wall\cite{exp_domainwall} call for a full
quantum treatment of the whole system. Edwards \emph{\ et.al.,}\cite%
{Edwards05} obtained the torques of spin valve in the empirical
tight-binding frame and Haney \emph{et.al.,}\cite{PMHaney} calculated the
torques in the similar structure with nonequilibrium Green's function (NEGF)
based on LCAO basis.

Both semiclassical and quantum mechanical study show that the STT is most
significant near the nonmagnet(NM)$|$FM interfaces in the spin valve. Up to
now, only a few studies have addressed the material dependence of spin
torque, which could be an important issue as the spin dependent transport is
greatly affected by the electronic structure in FM\cite%
{Maciej_decay,mixing_G_Turek}. Furthermore, previous studies focused on
ideal structure without considering the disorder at the FM$|$NM interface,
which should exist in the realistic spin valve\cite{flip06}.

The main aim of this paper is to formulate a method to calculate STT of a
noncollinear magnetized system within the first principles frame. Differing
from the previous Green function based work\cite{PMHaney}, we obtained the
complete scattering wave functions of the whole system\cite{Xia06}. The STT%
\cite{MDstiles02} is formulated in the tight-binding representation. Large
system such as domain wall can be well treated in this framework\cite{tang06}%
. We apply our formulism to the Co$|$Cu$|$FM$|$Cu spin valve system with
impurity scattering at the FM$|$NM interface. Our study shows that the STT
can penetrate deep into the ferromagnetic materials for Ni, which is quite
different from Co. It is also found that average torques are enhanced in the
presence of interfacial disorder.

This paper is organized as following. In Sec. II, we present the details of
the formalism for constructing the eigenmodes of the lead and computing the
STT in spin valve. Note that not only the transmission and reflection
coefficient are obtained but also the wave function in the scattering regime
is obtained explicitly. In Section III, the method is used to calculate the
conductance and STT in the systems of Co$|$Cu$|$FM$|$Cu(111), with FM is Co,
Ni and Ni$_{80}$Fe$_{20}$, respectively. The effect of interfacial disorder
is discussed. In Sec. IV, we summarize our results.

\section{Theoretical model}

\begin{figure}[tbp]
\includegraphics[width=8.6cm]{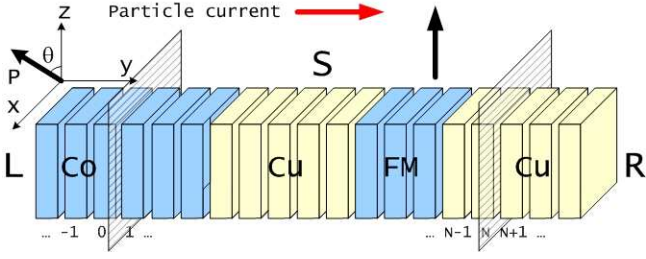}
\caption{(color online) Sketch of the configuration used for
current-induced switching. A scattering region is sandwiched by
left-($\mathbf{L}$) and right-hand($\mathbf{R}$)leads which have
translational symmetry and are partitioned into principle layers
perpendicular to the transport direction. The scattering region
contains $N$ principle layers but the structure and chemical
composition are in principle arbitrary. The switching layer FM can
be Co, Ni, Ni$_{80}$Fe$_{20}$.} \label{config}
\end{figure}

Let us focus on the spin transport and STT in the layered systems sketched
in Fig.\ref{config}. The scattering region $\mathbf{S}$, which is denoted by
the layer index $1\leq I\leq N$, is sandwiched by left$\left( \mathbf{L}%
\right) $ and right$\left( \mathbf{R} \right) $ leads. For this device,
there exists perfect lattice periodicity in the $X$-$Z$ plane. Particle
current flows along $Y$ axis. In scattering region no periodicity is assumed
along current direction. Here the atomic potentials were determined by the
tight-binding linearized muffin-tin-orbital (TB-LMTO) surface Green's
function (SGF) method\cite{I.Turek97book}. When combined with the coherent
potential approximation (CPA), this method allows the electronic structure,
charge, and spin densities of layered materials with substitutional disorder
to be calculated self-consistently with high efficiency. To model the
noncollinear system in the spin valve, the rigid potential approximation is
used. In this approximation, we rotate the potential of fixed magnet in spin
space to construct the relative angle between the polarization directions of
fixed magnet and free magnet, which is a good approximation as the two
magnets are spaced far enough by a Cu layer.

Following previous work\cite{Xia06}, we describe the theoretical frame
developed with wave-function matching (WFM) based on TB-LMTO basis for
studying the STT. In Sec. II A, we review the Hamiltonian and KKR equation
for a device with noncollinear magnetization. The equation of motion (EOM)
for layered system is extracted from KKR equation. In the Sec.II B, the
boundary conditions of the EOM are formulated in terms of the Bloch states
in the leads. In the Sec. II C, by solving the EOM in the scattering region
with embedding potentials of the two leads, we obtain the complete
scattering wave function of the scattering region. In the Sec. II D and E,
the particle current and spin current are formulated with those obtained
scattering wave function expanded in TB-LMTO basis.

\subsection{Hamiltonian and KKR equation}

For layered systems, atoms can always be grouped into principle layers
defined as to be so thick that the interactions between layers $I$ and $I\pm
2$ are negligible as shown in Fig.\ref{config}.

The EOM for $I$th principal layer can be written as
\begin{equation}
\mathbf{H}_{I,I-1}\mathbf{a}_{I-1}+\left( \mathbf{H}-E\right) _{II}\mathbf{a}%
_{I}+\mathbf{H}_{I,I+1}\mathbf{a}_{I+1}=0,  \label{eom}
\end{equation}%
where $E$ is always set to the Fermi energy $E_{F}$ for the transport
problem. Here, $\mathbf{a}_{I}$ is the a vector describing the amplitudes of
the $I$th layer in terms of the localized orbital basis $\left\vert \mathbf{R%
}L\zeta \right\rangle $, where $\mathbf{R}$ is the site index and $L$ can be
defined by $L\equiv (l,m)$. $l$ and $m$ are the azimuthal and magnetic
quantum numbers respectively. $\zeta =\uparrow \left( \downarrow \right) $
denotes that the basis is eigenstate in spin space, which is parallel
(antiparallel) to spin quantization axis.

To the first order approximation of the full LMTO\ Hamiltonian, a
short-range TB-LMTO\ Hamiltonian in the $\alpha $ representation\cite%
{J.Kudron00,Andersen85book} in the global coordinate system can be written
as
\begin{eqnarray}
\mathbf{H}_{\mathbf{R}L,\mathbf{R}^{\prime }L^{\prime }}^{\alpha } &=&U_{%
\mathbf{R}}\mathcal{\overline{C}}_{\mathbf{R}L}^{\alpha }U_{\mathbf{R}%
^{\prime }}^{\dagger }\delta _{\mathbf{R}^{\prime }L^{\prime }\mathbf{R}L}
\notag  \label{hamiltonian} \\
&&+[U_{\mathbf{R}}\left( \overline{\Delta }_{\mathbf{R}L}^{\alpha }\right) ^{%
\frac{1}{2}}U_{\mathbf{R}}^{\dag }S_{\mathbf{R}L,\mathbf{R}^{\prime
}L^{\prime }}^{\alpha }  \notag \\
&&\times U_{\mathbf{R}^{\prime }}\left( \overline{\Delta }_{\mathbf{R}%
^{\prime }L^{\prime }}^{\alpha }\right) ^{\frac{1}{2}}U_{\mathbf{R}^{\prime
}}^{\dagger }],
\end{eqnarray}%
where $\mathcal{\overline{C}}_{\mathbf{R}L}^{\alpha }$ and $\overline{\Delta
}_{\mathbf{R}L}^{\alpha }$ are $2\times 2$ potential parameter matrices
expanded in spin space and diagonal in the local coordinate system. The
unitary rotation matrix at site $\mathbf{R}$ can be defined by
\begin{equation}
U_{\mathbf{R}}\left( \theta _{\mathbf{R}},\varphi _{\mathbf{R}}\right) =%
\left[
\begin{array}{cc}
\cos \frac{\theta _{\mathbf{R}}}{2}e^{-i\frac{\varphi _{\mathbf{R}}}{2}} &
-\sin \frac{\theta _{\mathbf{R}}}{2}e^{-i\frac{\varphi _{\mathbf{R}}}{2}} \\
\sin \frac{\theta _{\mathbf{R}}}{2}e^{i\frac{\varphi _{\mathbf{R}}}{2}} &
\cos \frac{\theta _{\mathbf{R}}}{2}e^{i\frac{\varphi _{\mathbf{R}}}{2}}%
\end{array}%
\right] ,  \label{u}
\end{equation}%
where $\theta _{\mathbf{R}},\varphi _{\mathbf{R}}$ are the azimuth angles of
the local quantization axis. Screened structure constants $S_{\mathbf{R}L,%
\mathbf{R}^{\prime }L^{\prime }}^{\alpha }$ contain all information about
the structure, which are block diagonal in the spin space,
\begin{equation}
S_{\mathbf{R}L,\mathbf{R}^{\prime }L^{\prime }}^{\alpha }=\left[
\begin{array}{cc}
s_{\mathbf{R}L,\mathbf{R}^{\prime }L^{\prime }}^{\alpha } & 0 \\
0 & s_{\mathbf{R}L,\mathbf{R}^{\prime }L^{\prime }}^{\alpha }%
\end{array}
\right] .  \label{s}
\end{equation}
Note that $s_{\mathbf{R}L,\mathbf{R}^{\prime }L^{\prime }}^{\alpha }$ is
spin independent. The Hamiltonian of Eq.(\ref{hamiltonian}) yields
eigenvalues corrected to first order in $\left( E-E_{F}\right) $ and is
exact when we set $E=E_{F}$.

For a noncollinear magnetized system, the "tail-cancellation" condition
yields the KKR equation\cite{Andersen85book},
\begin{equation}
\underset{\mathbf{R}^{\prime },L^{\prime }}{\sum }\left( -U_{\mathbf{R}}%
\overline{P}_{\mathbf{R}L}^{\alpha }\left( E\right) U_{\mathbf{R}}^{\dag
}\delta _{\mathbf{RR}^{\prime }}\delta _{LL^{\prime }}-S_{\mathbf{R}L,%
\mathbf{R}^{\prime }L^{\prime }}^{\mathbf{\alpha }}\right) C_{\mathbf{R}%
^{\prime }L^{\prime }}=0,  \label{tail_cancellation}
\end{equation}%
where $\mathbf{C}_{\mathbf{R}L}=\left( \mathbf{C}_{\mathbf{R}L\uparrow },%
\mathbf{C}_{\mathbf{R}L\downarrow }\right) ^{T}$ has the relation to the
wave amplitude $\mathbf{a}_{\mathbf{R}L}$ of $L$ orbital at site $\mathbf{R}$
as $\mathbf{C}_{\mathbf{R}L}=U_{\mathbf{R}}\left( \overline{\Delta }_{%
\mathbf{R}L}^{\alpha }\right) ^{\frac{1}{2}}U_{\mathbf{R}}^{\dagger }\mathbf{%
a}_{\mathbf{R}L}$. $\overline{P}_{\mathbf{R}L}^{\alpha }\left( E\right) $ is
the screened potential function matrix and contains all information about
the atomic species at site $\mathbf{R}$ for calculating the electronic
structure. It is diagonal in the local coordinate system,
\begin{equation}
\overline{P}_{\mathbf{R}L}^{\alpha }\left( E\right) \equiv \left[
\begin{array}{cc}
\overline{p}_{\mathbf{R}L}^{\alpha ,\uparrow } & 0 \\
0 & \overline{p}_{\mathbf{R}L}^{\alpha ,\downarrow }%
\end{array}%
\right] ,  \label{p_func}
\end{equation}%
where $\overline{p}_{\mathbf{R}L}^{\alpha ,\uparrow \left( \downarrow
\right) }\equiv \left( E-\mathcal{\overline{C}}_{\mathbf{R}L}^{\alpha
,\uparrow \left( \downarrow \right) }\right) \left( \overline{\Delta }_{%
\mathbf{R}L}^{\alpha ,\uparrow \left( \downarrow \right) }\right) ^{-1}$ and
$E$ is set to $E_{F}$ for the transport problem we considered.

As there exists two-dimensional translational symmetry in the lateral plane,
the states along the transport direction can be characterized by a lateral
wave vector $\mathbf{k}_{\parallel }$ in the corresponding 2-dimensional
Brillouin zone (2D BZ). The screened KKR equation in the mixed
representation of $\mathbf{k}_{\parallel }$ can be expressed in terms of
principal layers as,
\begin{widetext}
\begin{equation}
-S_{I,I-1}^{\mathbf{k}_{\parallel }}\mathbf{C}_{I-1}\left(
\mathbf{k}_{\parallel}\right) +\left(
U_{I}\overline{P}_{I,I}\left( E_{F}\right) U_{I}^{\dag
}-S_{I,I}^{\mathbf{k}_{\parallel }}\right) \mathbf{C}_{I}\left(
\mathbf{k}_{\parallel}\right) -S_{I,I+1}^{\mathbf{k}_{\parallel
}}\mathbf{C}_{I+1}\left( \mathbf{k}_{\parallel}\right) =0,
\label{kkr}
\end{equation}
\end{widetext}where $\mathbf{C}_{I}\left( \mathbf{k}_{\parallel }\right) $
is the wave vector describing the wave function amplitudes of the $I$th
principal layer consisting of $h$ atom sites and has the dimension of $%
2\left( l_{\max }+1\right) ^{2}h=2M$. $\overline{P}_{I,I}$ is $2M\times 2M$
diagonal matrix. and $S_{I,I}^{\mathbf{k}_{\parallel }}$ is also $2M\times
2M $ matrices with its sub matrix $S_{\mathbf{R}L,\mathbf{R}^{\prime
}L^{\prime }}^{\mathbf{k}_{\Vert }}$ defined by
\begin{eqnarray}
&&S_{\mathbf{R}L,\mathbf{R}^{\prime }L^{\prime }}^{\mathbf{k}_{\Vert }}
\notag \\
&=&\underset{\mathbf{T}_{\Vert }}{\sum }\exp \left[ i\mathbf{k}_{\Vert
}\cdot \mathbf{T}_{\parallel }\right] S_{\mathbf{R}L,\left( \mathbf{R}%
^{\prime }+\mathbf{T}_{\parallel }\right) L^{\prime }}^{\alpha }\text{ \ }%
\begin{array}{c}
\mathbf{R}\in I \\
\mathbf{R}^{\prime }\mathbf{+T}_{\parallel }\in I^{\prime }%
\end{array}%
,  \label{structure_k}
\end{eqnarray}%
where $I$ and $I^{\prime }$ are layer index and $\mathbf{T}_{\parallel }$ is
2-dimensional translational vector in the plane of principal layer.

Note that Eq.(\ref{kkr}) is the EOM by analogy with Eq.(\ref{eom}). We will
solve it for a given energy $E_{F}$ of electrons to obtain the wave function
of the scattering state. The reference to $\mathbf{k}_{\parallel}$ and $%
E_{F} $ in the formulism will be suppressed in the following two parts Sec.
II B and Sec. II C.

\subsection{Eigenmodes of the leads}

For the scattering problem, far enough away from the scattering region the
wave function can be expressed rigorously with asymptotic forms in terms of
reflection and transmission coefficients and Bloch states in the leads. As
the wave function should satisfy Bloch's theorem in a periodic potential, we
set $\mathbf{\overline{C}}_{n}\mathbf{=}\lambda ^{n}\mathbf{\overline{C}}%
_{0} $. In local coordinate system, the EOM in lead becomes
\begin{eqnarray}
&&\left(
\begin{array}{cc}
S_{0,1}^{-1}\left( \overline{P}_{00}-S_{0,0}\right) & -S_{0,1}^{-1}S_{1,0}
\\
1 & 0%
\end{array}%
\right) \left(
\begin{array}{c}
\mathbf{\overline{C}}_{0} \\
\mathbf{\overline{C}}_{-1}%
\end{array}%
\right)  \notag \\
&=&\lambda \left(
\begin{array}{c}
\mathbf{\overline{C}}_{0} \\
\mathbf{\overline{C}}_{-1}%
\end{array}%
\right).  \label{lead_u}
\end{eqnarray}%
Details for solving Bloch states $\mathbf{\overline{C}}_{0}$ can be found in
Ref.[\onlinecite{Xia06}]. To overcome the numerical difficult of the spin
degeneracy in NM lead and reduce the calculation efforts, we solve the EOM
in the leads in local coordinate system for each spin separately. In global
coordinate system, Bloch states can be obtained after an unitary
transformation. For the amplitude of 0th layer, we have $\mathbf{C}_{0}=U_{0}%
\mathbf{\overline{C}} _{0}$.

The propagating states and evanescent states can be identified and sorted
into right-going$\left( +\right) $ or left-going $\left( -\right) $. Letting
$\mathbf{\overline{w}}_{\mu }^{\uparrow \left( \downarrow \right) }\left(
\pm\right) $ denotes the solutions of $\mathbf{\overline{C}}$ corresponding
to eigenvalue $\lambda _{\mu }\left( \pm\right) $, where $\uparrow $($%
\downarrow $) denotes the eigenstate parallel (antiparallel) to the local
spin quantization direction. Constructing the matrix $\mathbf{W}\left( \pm
\right) $ as

\begin{eqnarray}
\mathbf{W}\left( \pm \right) &=&U_{0}\mathbf{\overline{W}}\left( \pm \right)
\notag \\
&\equiv & U_{0}[\overline{\mathbf{w}}_{1}^{\uparrow }\left( \pm \right)
,\cdots ,\overline{\mathbf{w}}_{M}^{\uparrow }\left( \pm \right) ,  \notag \\
&&\text{ \ \ \ \ \ }\overline{\mathbf{w}}_{1}^{\downarrow }\left( \pm
\right) ,\cdots ,\overline{\mathbf{w}}_{M}^{\downarrow }\left( \pm \right) ].
\label{u_b}
\end{eqnarray}

Following Ando\cite{T.Ando91}, define the Bloch factor as

\begin{equation}
F\left( \pm \right) \equiv \mathbf{W}\left( \pm \right) \mathbf{\Lambda }%
\left( \pm \right) \mathbf{W}^{-1}\left( \pm \right) .  \label{bloch_ma}
\end{equation}%
where $\mathbf{\Lambda }\left( \pm \right) $ is a diagonal matrix with the
diagonal elements given by $\left[ \lambda _{1}^{\uparrow }\left( \pm
\right) ,\cdots ,\lambda _{M}^{\uparrow }\left( \pm \right) ,\lambda
_{1}^{\downarrow }\left( \pm \right) ,\cdots ,\lambda _{M}^{\downarrow
}\left( \pm \right) \right] $. In local coordinate system, we have the
relation $\mathbf{\overline{C}}_{I}\left( \pm \right) =\overline{F}%
^{I-J}\left( \pm \right) \mathbf{\overline{C}}_{J}\left( \pm \right) $\cite%
{Xia06}.It is easy to proof that the Bloch factor defined above satisfies
the Bloch relation in global coordinate system
\begin{equation}
\mathbf{C}_{I}\left( \pm \right) =F\left( \pm \right) ^{I-J}\mathbf{C}%
_{J}\left( \pm \right) .  \label{c_relation}
\end{equation}
Bloch factors matrix $F\left( \pm \right) $ relates the wave amplitude in
the $I$th layer to that in the $J$th layer for a state in the lead.

\subsection{Scattering problem}

The equations of motion with open boundary conditions for a device usually
contain infinite number of equations. By incorporating the boundary
conditions in the leads, the scattering problem can be reduced to a set of
coupled linear equations with finite number of equations\cite{Xia06}.

For an electron coming from the left lead, Eq.(\ref{kkr}) for $I=0$ can be
rewritten as

\begin{eqnarray}
&&\left( U_{0}\overline{P}_{0,0}U_{0}^{\dag }-\tilde{S}_{0,0}\right) \mathbf{%
C}_{0}-S_{0,1}\mathbf{C}_{1}  \notag \\
&=&S_{0,-1}\left[ F_{L}^{-1}\left( +\right) -F_{L}^{-1}\left( -\right) %
\right] \mathbf{C}_{0}\left( +\right) ,
\end{eqnarray}%
where $L$ denotes the left lead and with $\tilde{S}_{0,0}\equiv
S_{0,0}+S_{0,-1}F_{L}^{-1}\left( -\right) $. The $S_{0,-1}F_{L}^{-1}\left(
-\right) $ is the embedding potential for the left lead.

In the right lead, only right-going waves exist in the $\left( N+1\right)$th
layer. The EOM for $I=N+1$ is
\begin{equation}
\left( U_{N+1}\overline{P}_{N+1,N+1}U_{N+1}^{\dag }-\tilde{S}%
_{N+1,N+1}\right) \mathbf{C}_{N+1}-S_{N+1,N}\mathbf{C}_{N}=0,
\end{equation}%
where $\tilde{S}_{N+1,N+1}=S_{N+1,N+1}+S_{N+1,N+2}F_{R}\left( +\right) $ and
$S_{N+1,N+2}F_{R}\left( +\right) $ is the embedding potential for the right
lead.

Making use of the lead boundary conditions for $0$th and $\left( N+1\right) $
layer, the scattering wave function can be found as

$\left(
\begin{array}{c}
\mathbf{C}_{0} \\
\mathbf{C}_{1} \\
\mathbf{C}_{2} \\
\mathbf{\vdots } \\
\mathbf{C}_{N} \\
\mathbf{C}_{N+1}%
\end{array}%
\right) =\left( U\mathbf{\overline{P}}U^{\dag }\mathbf{-\tilde{S}}\right)
^{-1}$

\begin{equation}
\text{ \ \ \ \ \ \ \ \ }\times \left(
\begin{array}{c}
S_{1,-1}\left[ F_{L}^{-1}\left( +\right) -F_{L}^{-1}\left( -\right) \right]
\mathbf{C}_{0}\left( +\right) \\
0 \\
\vdots \\
0 \\
0%
\end{array}%
\right),  \label{coeff}
\end{equation}%
where $\mathbf{\tilde{S}}$ is of block tridiagonal matrix containing $%
S_{I,J} $ except the $\tilde{S}_{0,0}$ and $\tilde{S}_{N+1,N+1}$ are defined
as above. The spin polarization direction at different sites can be
incorporated by the unitary rotation $U$ at corresponding site.

To obtain the scattering state, we need to specify an incoming state $%
\mathbf{C}_{0}\left( +\right) $ at the right side of Eq.(\ref{coeff}). This
can be achieved by introducing the right going eigenmodes of left lead as
the incoming states by setting $\mathbf{C}_{0}\left( +\right)$ to be $%
\mathbf{w}_{\lambda }\left( +\right) $, where $\mathbf{w}_{\lambda }\left(
+\right) $ should be renormalized so as to carry an unit flux. Each $\mathbf{%
w}_{\lambda }\left( +\right) $ corresponds to a scattering state in device.

The amplitude of layers from $0$ to $N+1$ solved from Eq.(\ref{coeff})
serves for computing the particle current and spin current. Also, the
scattering matrix can be obtained\cite{Xia06}.

\subsection{Particle Current}

Let us consider the particle current operator of a quasi one-dimensional TB
model for a special $\mathbf{k}_{\parallel }$ vector at $E=E_{F}$. The
MTO-basis functions $\left\vert \mathbf{R}L\zeta ^{\mathbf{k}_{\parallel
}}\right\rangle $ are obtained from the Bloch sum of the particle waves:%
\begin{equation}
\left\vert \mathbf{R}L\zeta ^{\mathbf{k}_{\parallel }}\right\rangle =%
\underset{T_{\parallel }}{\sum }e^{i\mathbf{k}_{\parallel }\cdot \mathbf{T}%
_{\parallel }}\left\vert \mathbf{R+T}_{\parallel },L\zeta ^{\alpha
}\right\rangle .  \label{orbital_tran}
\end{equation}%
So the density operator at $\mathbf{R}$ site in the mixed representation for
a special $\mathbf{k}_{\parallel }$ vectors can be defined by
\begin{equation}
\mathbf{\hat{\rho}}_{\mathbf{R}}^{\mathbf{k}_{\parallel }}\equiv \underset{%
L\zeta }{\sum }\left\vert \mathbf{R}L\zeta ^{\mathbf{k}_{\parallel
}}\right\rangle \left\langle \mathbf{R}L\zeta ^{\mathbf{k}_{\parallel
}}\right\vert .  \label{density_k}
\end{equation}

Neglecting the electron motion inside the atomic cells, the velocity
operators can be expressed by the intersite hopping\cite{Turek02} and will
give the total current for subspace. The velocity (current) operator can be
defined by%
\begin{equation}
\mathbf{\hat{V}}=\frac{1}{i\hbar }\left[ \mathbf{\hat{X},\hat{H}}\right] ,
\label{v_d}
\end{equation}%
where $\mathbf{\hat{X}}$ is the coordinate operator, which can be
represented in TB model by a diagonal matrix $\mathbf{\hat{X}}_{\mathbf{R}L,%
\mathbf{R}^{\prime }L^{\prime }}=\mathbf{X}_{\mathbf{R}}\delta _{\mathbf{RR}%
^{\prime }}\delta _{LL^{\prime }}$ \cite{Turek02}.

With aid of Eq.(\ref{v_d}), the current operator $\mathbf{\hat{J}}_{\mathbf{R%
}^{\prime }\mathbf{R}}^{\mathbf{k}_{\parallel }}$ from $\mathbf{R}^{\prime }$%
th to $\mathbf{R}$th site ($\mathbf{R}\neq \mathbf{R}^{\prime }$) can be
written as
\begin{equation}
\mathbf{\hat{J}}_{\mathbf{R}^{\prime }\mathbf{R}}\left( \mathbf{k}%
_{\parallel }\right) =\underset{LL^{\prime }}{\sum }\frac{1}{i\hslash }\left[
\mathbf{\hat{H}}_{\mathbf{R}L,\mathbf{R}^{\prime }L^{\prime }}^{\mathbf{k}%
_{\parallel }}-h.c.\right] .  \label{j_operator1}
\end{equation}%
where $\mathbf{\hat{H}}_{\mathbf{R}L\zeta ,\mathbf{R}^{\prime }L^{\prime
}\zeta ^{\prime }}^{\mathbf{k}_{\parallel }}=\left\vert \mathbf{R}L\zeta ^{%
\mathbf{k}_{\parallel }}\right\rangle \mathbf{H}_{\mathbf{R}L\zeta ,\mathbf{R%
}^{\prime }L^{\prime }\zeta ^{\prime }}^{\mathbf{k}_{\parallel
}}\left\langle \mathbf{R}^{\prime }L^{\prime }\zeta ^{\prime \mathbf{k}%
_{\parallel }}\right\vert $ and $\mathbf{H}_{\mathbf{R}L,\mathbf{R}^{\prime
}L^{\prime }}^{\mathbf{k}_{\parallel }}$ is the Hamiltonian matrix in spin
space, which has relation with Eq.(\ref{hamiltonian}) as%
\begin{equation}
\mathbf{H}_{\mathbf{R}L,\mathbf{R}^{\prime }L^{\prime }}^{\mathbf{k}%
_{\parallel }}=\underset{\mathbf{T}_{\parallel }}{\sum }\exp \left[ i\mathbf{%
k}_{\parallel }\cdot \mathbf{T}_{\parallel }\right] \mathbf{H}_{\mathbf{R}L%
\mathbf{,}\left( \mathbf{R}^{\prime }+\mathbf{T}_{\parallel }\right)
L^{\prime }}^{\alpha }.  \label{hami_k}
\end{equation}

The expectation value of operator $\mathbf{\hat{A}}$ is $\left\langle
\mathbf{\hat{A}}\right\rangle \equiv \left\langle \Psi \left\vert \mathbf{%
\hat{A}}\right\vert \Psi \right\rangle $. The particle current can be
expressed as
\begin{equation}
\left\langle \mathbf{\hat{J}}_{\mathbf{R}^{\prime }\mathbf{R}}\left( \mathbf{%
k}_{\parallel }\right) \right\rangle =\underset{LL^{\prime }}{\sum }\frac{1}{%
i\hslash }[\mathbf{a}_{\mathbf{R}L}^{\dagger }\left( \mathbf{k}_{\parallel
}\right) \mathbf{H}_{\mathbf{R}L,\mathbf{R}^{\prime }L^{\prime }}^{\mathbf{k}%
_{\parallel }}\mathbf{a}_{\mathbf{R}^{\prime }L^{\prime }}\left( \mathbf{k}%
_{\parallel }\right) -h.c.],  \label{j_k_a}
\end{equation}%
where where $\mathbf{a}_{\mathbf{R}L}\left( \mathbf{k}_{\parallel }\right)
=\left( \mathbf{a}_{\mathbf{R}L\uparrow }\left( \mathbf{k}_{\parallel
}\right) ,\mathbf{a}_{\mathbf{R}L\downarrow }\left( \mathbf{k}_{\parallel
}\right) \right) ^{T}$, and $\mathbf{a}_{\mathbf{R}L\zeta }\left( \mathbf{k}%
_{\parallel }\right) =\left\langle \mathbf{R}L\zeta ^{\mathbf{k}_{\parallel
}}|\Psi \right\rangle $. $\mathbf{a}_{\mathbf{R}L}\left( \mathbf{k}%
_{\parallel }\right) $ has the relation with $\mathbf{C}_{\mathbf{R}L}\left(
\mathbf{k}_{\parallel }\right) $ as follow%
\begin{equation}
\mathbf{a}_{\mathbf{R}L}\left( \mathbf{k}_{\parallel }\right) =U_{\mathbf{R}%
}\left( \overline{\Delta }_{\mathbf{R}L}^{\alpha }\right) ^{-\frac{1}{2}}U_{%
\mathbf{R}}^{\dagger }\mathbf{C}_{\mathbf{R}L}\left( \mathbf{k}_{\parallel
}\right) .  \label{a_c}
\end{equation}

The $\mathbf{C}_{\mathbf{R}L}\left( \mathbf{k}_\parallel\right)$ can be
obtained by Eq.(\ref{coeff}) for a given $k_\parallel$. Within the MTO
formulism, the current can also be expressed with structure constants matrix
as in Ref.\onlinecite{Xia06}
\begin{equation}
\left\langle \mathbf{\hat{J}}_{\mathbf{R}^{\prime }\mathbf{R}}\left( \mathbf{%
k}_{\parallel}\right) \right\rangle =\underset{LL^{\prime }}{\sum }\frac{1}{%
i\hslash }[ \mathbf{C}_{\mathbf{R}L}^{\dagger }\left( \mathbf{k}%
_{\parallel}\right) S_{\mathbf{R}L,\mathbf{R}^{\prime }L^{\prime }}^{\mathbf{%
k}_{\parallel}}\mathbf{C}_{\mathbf{R}^{\prime }L^{\prime }}\left( \mathbf{k}%
_{\parallel}\right)-h.c.].  \label{j_lmto}
\end{equation}

\begin{figure}[tbp]
\includegraphics[width=6cm]{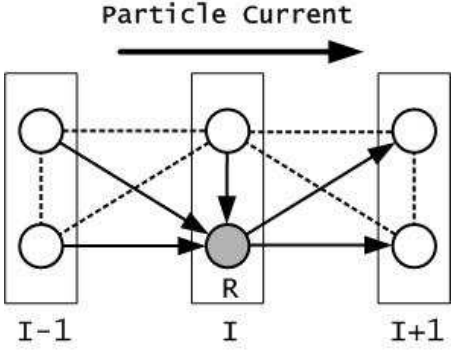}\newline
\caption{Illustration of incoming current and outcoming current for the $%
\mathbf{R}$th site. Assuming the particle current comes from $I-1$th layer
to $I+1$th layer. Arrow lines denote the current related to $\mathbf{R}$th
site and dot lines denote the coupling between sites irrelevant to $\mathbf{R%
}$th site. } \label{c2}
\end{figure}

The continuity equation of particle current at $\mathbf{R}$ site in the $I $%
th principle layer reads
\begin{eqnarray}
&&\underset{\mathbf{R}^{\prime }\in I-1,I}{\sum }\left\langle \mathbf{\hat{J}%
}_{\mathbf{R}^{\prime },\mathbf{R}}\left( \mathbf{k}_{\parallel}\right)
\right\rangle -\underset{\mathbf{R}^{\prime }\in I,I+1}{\sum }\left\langle
\mathbf{\hat{J}}_{\mathbf{R},\mathbf{R}^{\prime }}\left( \mathbf{k}%
_{\parallel}\right) \right\rangle  \notag \\
&=&\frac{\mathbf{d}\left\langle \mathbf{\hat{\rho}}_{\mathbf{R}}^{\mathbf{k}%
_{\parallel}}\right\rangle }{\mathbf{d}t}\text{ \ },  \label{j_conti}
\end{eqnarray}
where the first term at the left side of Eq.(\ref{j_conti}) is the incoming
current to the $\mathbf{R}$ site and the second term is outgoing current
from this site.

As shown in Fig.(\ref{c2}), the current is assumed to flow from $I-1$th
layer to $I+1$th layer. Considering the $\mathbf{R}$ site, the incoming
current is composed of current from the sites in the $I-1$th layer and the
sites ahead $\mathbf{R}$ site (relative to transport direction) in $I$th
layer. If there exist other atoms in the same plane of $\mathbf{R}$(see Fig.(%
\ref{c2})), the current from those atoms also are considered as the
component of incoming current to $\mathbf{R}$ site. The outgoing current is
composed of current to the sites in $I+1$th layer and those sites behinds $%
\mathbf{R}$ site in $I$th layer. Note that treating the current between
atoms in the same plane as the incoming current or outgoing current will not
result in any physical consequence. Careful check has been carried out that
the particle current conservation law can be satisfied atom by atom and
layer by layer. For the scattering states we calculated, the right side of
the Eq.(\ref{j_conti} ) is zero.

In the linear response regime, the particle current under a small bias $%
V_{b} $ at zero temperature can be expressed as\cite{Datta95},
\begin{equation}
\mathbf{J}_{\mathbf{RR}^{\prime }}=\frac{e}{h}\frac{1}{N_{\parallel }}%
\underset{\mathbf{k}_{\parallel }}{\sum }\left\langle \mathbf{\hat{J}}_{%
\mathbf{RR}^{\prime }}\left( \mathbf{k}_{\parallel }\right) \right\rangle
V_{b},  \label{j_linear}
\end{equation}%
where the bias is given by the difference between the electrochemical
potentials of the two leads as $eV_{b}=\mu _{L}-\mu _{R}$, and $N_{\parallel
}$ is the number of $\mathbf{k}_{\parallel }$ points in 2D BZ.

\subsection{Spin current and spin torque}

The spin current is defined similar to the particle current in Section II.D.
Considering a quasi one-dimensional TB mode for a special $\mathbf{k}%
_{\parallel }$ vector, the spin density operator at site $\mathbf{R}$ can be
defined as
\begin{equation}
\mathbf{\hat{S}}_{\mathbf{R}}^{\mathbf{k}_{\parallel }}\equiv \underset{%
L\zeta }{\sum }\left\vert \mathbf{R}L\zeta ^{\mathbf{k}_{\parallel
}}\right\rangle \mathbf{\hat{\sigma}}\left\langle \mathbf{R}L\zeta ^{\mathbf{%
k}_{\parallel }}\right\vert ,  \label{s_density}
\end{equation}%
where $\mathbf{\hat{\sigma}}$ is $2\times 2$ Pauli spin matrix. The spin
current operator generally can be defined as
\begin{equation}
\hat{\mathcal{J}}\equiv \frac{1}{2}\left[ \mathbf{\hat{\sigma}}\otimes
\mathbf{\hat{V}}+\mathbf{\hat{V}}\otimes \mathbf{\hat{\sigma}}\right] .
\label{js_d}
\end{equation}
note that $\hat{\mathcal{J}}$ is a tensor. For spin current between $\mathbf{%
R}$th and $\mathbf{R}^{\prime }$th site ($\mathbf{R}\neq \mathbf{R}^{\prime
} $), we could project $\hat{\mathcal{J}}$ along the direction vector $%
\mathbf{x}_{\mathbf{R},\mathbf{R}^{\prime }}$ in real space as $\hat{%
\mathcal{J}}\cdot \mathbf{x}_{\mathbf{R},\mathbf{R}^{\prime }}$. Then the
spin current operator $\hat{\mathcal{J}}_{\mathbf{R}^{\prime },\mathbf{R}%
}\left( \mathbf{k}_{\parallel }\right) $ from $\mathbf{R}^{\prime }$th to $%
\mathbf{R}$th site ($\mathbf{R}\neq \mathbf{R}^{\prime }$) can be written as

\begin{equation}
\hat{\mathcal{J}}_{\mathbf{R}^{\prime },\mathbf{R}}\left( \mathbf{k}%
_{\parallel }\right) =\underset{LL^{\prime }}{\sum }\frac{1}{2i\hslash }[%
\mathbf{\hat{\sigma}\hat{H}}_{\mathbf{R}L,\mathbf{R}^{\prime }L^{\prime }}^{%
\mathbf{k}_{\parallel }}+\mathbf{\hat{H}}_{\mathbf{R}L,\mathbf{R}^{\prime
}L^{\prime }}^{\mathbf{k}_{\parallel }}\mathbf{\hat{\sigma}}-h.c.].
\label{js_operator}
\end{equation}
where $\hat{\mathcal{J}}_{\mathbf{R}^{\prime },\mathbf{R}}\left( \mathbf{k}%
_{\parallel }\right) $ is a vector only in spin space.

For a specific state $|\Psi \rangle $, the expectation value is
\begin{widetext}
\begin{equation}
\left\langle \hat{\mathcal{J}}_{\mathbf{R}^{\prime
},\mathbf{R}}\left(
\mathbf{k}_{\parallel }\right) \right\rangle =\underset{LL^{\prime }}{\sum }%
\frac{1}{2i\hslash }\left[ \mathbf{a}_{\mathbf{R}L}^{\dag }\left( \mathbf{k}%
_{\parallel }\right) \mathbf{\hat{\sigma}H_{\mathbf{R}L,\mathbf{R}^{\prime
}L^{\prime }}^{\mathbf{k}_{\parallel }}a}_{\mathbf{R}^{\prime }L^{\prime
}}\left( \mathbf{k}_{\parallel }\right) +\mathbf{a}_{\mathbf{R}L}^{\dag
}\left( \mathbf{k}_{\parallel }\right) \mathbf{H}_{\mathbf{R}L,\mathbf{R}%
^{\prime }L^{\prime }}^{\mathbf{k}_{\parallel }}\mathbf{\hat{\sigma}a}_{%
\mathbf{R}^{\prime }L^{\prime }}\left( \mathbf{k}_{\parallel }\right) -h.c.%
\right] .  \label{js_average}
\end{equation}
\end{widetext}

The STT $\left\langle \mathbf{\hat{T}}_{\mathbf{R}}^{s}\left( \mathbf{k}%
_{\parallel }\right) \right\rangle $ can be defined as the difference of the
incoming spin current and outgoing spin current of $\mathbf{R}$ site in the $%
I$th principal layer:
\begin{eqnarray}
&&\left\langle \mathbf{\hat{T}}_{\mathbf{R}}^{s}\left( \mathbf{k}_{\parallel
}\right) \right\rangle  \notag  \label{t_average} \\
&=&\underset{\mathbf{R}^{\prime }\in I-1,I}{\sum }\left\langle \hat{\mathcal{%
J}}_{\mathbf{R}^{\prime },\mathbf{R}}^{s}\left( \mathbf{k}_{\parallel
}\right) \right\rangle -\underset{\mathbf{R}^{\prime }\in I,I+1}{\sum }%
\left\langle \hat{\mathcal{J}}_{\mathbf{R},\mathbf{R}^{\prime }}^{s}\left(
\mathbf{k}_{\parallel }\right) \right\rangle .  \notag \\
&&
\end{eqnarray}
where the superscript $s$ is used to denote the incoming state is parallel
or antiparallel to the local spin quantization axis of injection lead, which
is very helpful, e.g. in ferromagnet we can distinguish the contribution to
the total torques from the majority spin or minority spin. Such definition
consists with those in Ref.[\onlinecite{PMHaney}], where analytic analysis
shows that for STT defined in this way equals to the exchange torques acted
on the injected spin defined in Eq.(\ref{s_density}) with only a sign
difference. After summation over 2D BZ, spin torque acted on $\mathbf{R}$ th
atom can be expressed as
\begin{equation}
\mathbf{T}_{\mathbf{R}}=\left( \frac{\hbar }{2}\right) \frac{e}{h}\frac{1}{%
N_{\parallel }}\underset{s,\mathbf{k}_{\parallel }}{\sum }\left\langle
\mathbf{\hat{T}}_{\mathbf{R}}^{s}\left( \mathbf{k}_{\parallel }\right)
\right\rangle V_{b},  \label{t_linear}
\end{equation}%
where the bias is given by the difference between the electrochemical
potentials of the two leads as $eV_{b}=\mu _{L}-\mu _{R}$.

\section{Spin transfer torques in Co/Cu/FM/Cu (111) Spin valves}

\subsection{Ordered interfaces}

A spin valve of Co$|$Cu$|$FM$|$Cu as shown in Fig.\ref{config} is used as an
example to illustrate our method. The left lead consists of semi-infinite Co
with the polarization direction $\theta $ (see Fig.\ref{config}). Cu spacer
of $9$ monolayer (ML) is located between fixed magnet Co and free magnet FM.
The free magnet contains $d$ ML, which could be Co, Ni, or Ni$_{80}$Fe$_{20}$
in this study. The lattice constants is assumed to be uniform in the whole
spin valve, that is, $a_{Cu}=a_{FM}=3.54\mathring{A}$ and the transport is
along \emph{fcc}[111]. With $spd$ -basis, exchang-correlation potential is
calculated and parameterized by Vosko-Wilk-Nusair \cite{vosko}. Our
calculation gives the magnetic moments as $1.64\mu _{B}/$Co atom, $2.60\mu
_{B}/$Fe atom and $0.62\mu _{B}/$Ni atom. For the calculation of transport,
total 90000 $\mathbf{k}_{\parallel}$ points in 2D BZ are summed.
\begin{figure}[tbp]
\includegraphics[width=8cm]{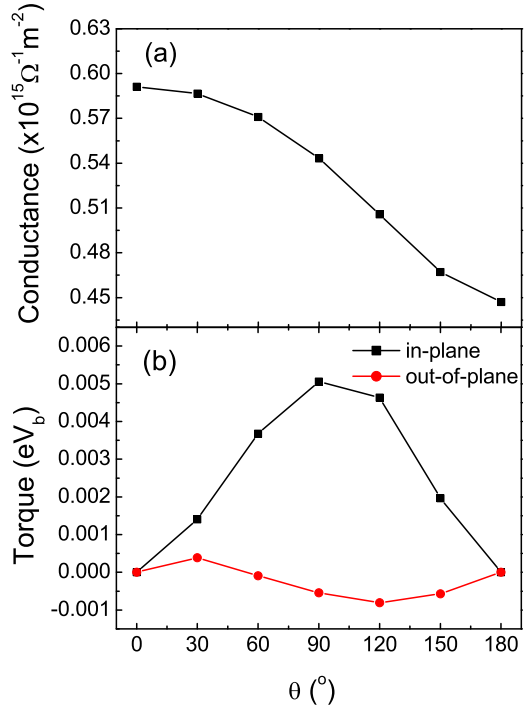}
\caption{(color online) (a)The total conductance of Co($\protect\theta $)$|$%
Cu(9ML)$|$Co(15ML)$|$Cu versus polarization direction
$\protect\theta $ of fixed magnet Co. (b) The angular dependence of
total spin torques on free magnet Co, where the electron current
flow from the fixed magnet to the free magnet. } \label{cocucond}
\end{figure}

Firstly, we present the angular dependence of total conductance $G(\theta)$
of the spin valve with the free magnet to be 15ML Co in Fig.\ref{cocucond}
(a). The monotonic decrease with increase of $\theta$ is consistent with the
previous \emph{ab.initio} results\cite{PMHaney}. Giant magnetoresistance
(GMR) can be defined as $GMR\equiv \frac{G(0^{o})-G(180^{o})}{G(180^{o})}%
100\%$, which is $24\%$ in this case. With electron current flowing from the
fixed magnet to the free magnet, Fig.\ref{cocucond} (b) gives the angular
dependence of total spin torques on the free magnet Co, which restore the
line shape of spin torques obtained in previous work\cite{PMHaney}. With the
drive of the in-plane torque, magnetization of free layer is going to
parallel to that of fixed layer. Due to the breakdown of time reverse
symmetry for spin current, if the direction of electron current is reversed,
the in-plane torques is going to drive free layer to antiparallel to fixed
layer. Such phenomenon is exactly the current induced switching of
magnetization observed in spin valve.

\begin{figure}[tbp]
\includegraphics[width=8cm, bb=9 9 191 227]{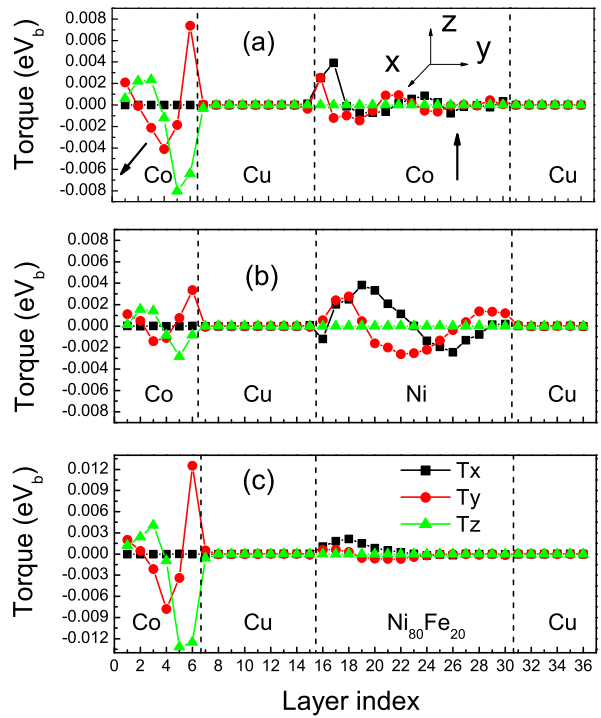}
\caption{(color online) The layer dependence of STT on interfacial
unit cell in the spin valve Co($\protect\theta =90$)$|$Cu$|$FM$|$Cu
, where the free magnet FM are Co, Ni, Ni$_{80}$Fe$_{20}$
respectively.} \label{torque3}
\end{figure}

\emph{Layer resolved Spin Torque:} The layer resolved STT contains the
information about whether the spin angle moment is absorbed near the
interface or not. Fig.\ref{torque3} gives the comparison of the layer
dependence of STT in the spin valves with three different free magnet. Here
the polarization of the fixed magnet Co is set to $\theta =90^{o}$ without
lost of generality. In our frame, $\mathbf{T}_{x}$ corresponds to the
in-plane torques and $\mathbf{T}_{y}$ corresponds to the out-of-plane
torque. The decay and oscillation of the STT are greatly different among
those materials we studied. When free magnet is Co as shown in Fig.\ref%
{torque3}(a), our result almost reproduces the previous result\cite%
{Edwards05,PMHaney}. The fast decay of the STT indicated the surface atoms
absorbed most of the spin angle moment as the current passes by.

However, when Ni serves as the free magnet as shown in Fig\ref{torque3}(b),
the maximum torques is not on the surface atom and the decay is very slow
with much longer oscillation. This observation is quite different with our
previous knowledge\cite{Maciej_decay}. The similar behavior is also found in
Fig\ref{torque3}(c) as Ni$_{80}$Fe$_{20}$ is free magnet, the oscillation
looks like that in Ni, but decaying faster. Due to the lack of obvious
oscillation, the total in-plane torques on Ni$_{80}$Fe$_{20}$ ($7.7\times
10^{-3}eV_{b}$) is greater than that on Co ($5.0\times 10^{-3}eV_{b}$) and
that on Ni ($6.3\times 10^{-3}eV_{b}$).

\begin{figure}[tbp]
\includegraphics[width=8.6cm]{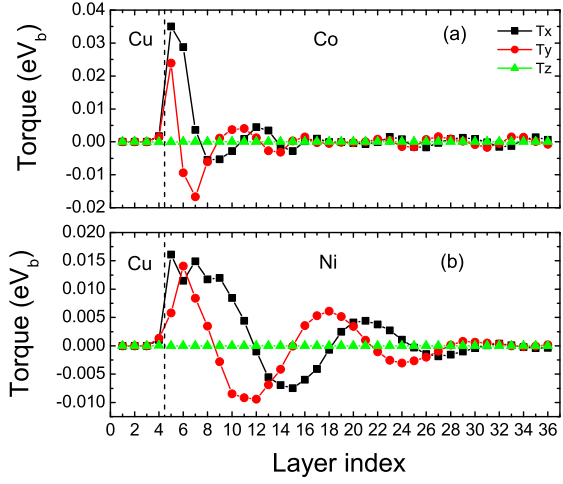}
\caption{(color online) The layer dependence of STT on interfacial
unit cell for the spin injection setup of single interface. (a)
Cu$|$Co (b)Cu$|$Ni.} \label{coni}
\end{figure}

The layer resolved STT shown in Fig.\ref{torque3} could be affected by the
multiple scattering between the two interfaces with Cu. To remove multiple
scattering effect on the torque, we perform the calculation for single
interfaces of Cu($90^{o}$)$|$Co and Cu($90^{o}$)$|$Ni, with 100\% polarized
electrons injected from Cu side. Here Cu($90^{o}$) indicates the
polarization direction of the injected electrons. The results are shown in
Fig.\ref{coni}. For both interfaces, the maximum torques exists on the
surface atoms and the oscillation spectra in ferromagnet become smoother and
clear. Still the oscillation behavior strongly depends on the materials. The
STT exists only near the Cu$|$Co interface, while the STT penetrates deep
into the Ni for Cu$|$Ni interface. Due to the long penetration length, the
multiple scattering between two Cu$|$Ni interface does appear in Fig.\ref%
{torque3}(b).

\emph{Simple Model for Spin Torques in FM:}For the layered system such as
spin valve, the incoming state of the injection lead can be labelled by $%
\mathbf{k}_{\parallel}$ in 2D BZ. Generally, these states will be coupled to
the propagating states and evanescent states of another side of the
injection interface. The STT can be expressed as

\begin{figure}[tbp]
\includegraphics[width=8.6cm]{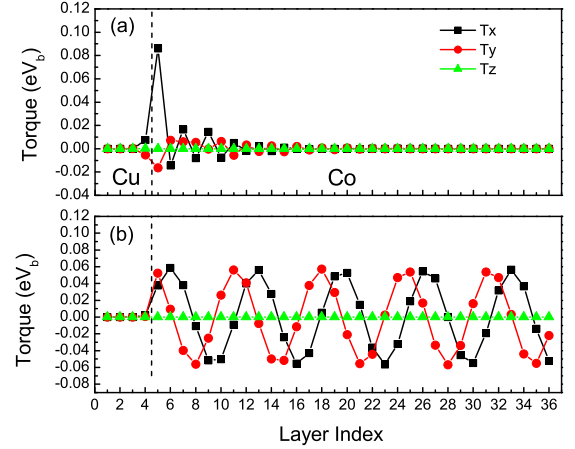}
\caption{(color online) (a) and (b) The layer dependence of STT on
interfacial unit cell when the spin injected through a single
Cu$|$Co interface for different $\mathbf{k}_{\parallel}$ points in
2D BZ. } \label{decay_bz}
\end{figure}

\begin{equation}  \label{torque_pression}
\Gamma\varpropto\sum_{\mu,\nu } C_{\mu\nu} e^{i[(k_{\mu }^{\downarrow
}-k_{\nu }^{\uparrow })x+\varphi_{\mu\nu}]}+\Im_{decay}(x),
\end{equation}
where first term denotes the contribution from the propagating states\cite%
{MDstiles02} and $k_{\mu }^{\downarrow }-k_{\nu}^{\uparrow }$ gives the
spatial precession frequency $\bigtriangleup k_{\mu \nu }$. The contribution
from the propagating states should oscillate as function of position and
will not decay as shown in Fig.\ref{decay_bz}(b). However, the frequency
could be quite different as $\mathbf{k}_{\parallel}$ runs over the 2DBZ, so
the final contribution will decay after summation over 2DBZ. The second term
of Eq. (\ref{torque_pression}), $\Im_{decay}(x)$, is the contribution from
the evanescent states. As we have known that no particle current can be
carried by evanescent state, however, such states do give effect on the spin
current and also on the STT. Evanescent states do contribute to spin torques
and should responds for the initio decay of the STT in the system as Co$|$Cu$%
|$Co$|$Cu, as shown in Fig.\ref{decay_bz}(a) where the evanescent state
dominates.

Two reasons could account for the decay of the STT away from the interface.
(i) Vanishing of the evanescent states' contribution. For Cu$|$Co, our
calculation shows that this part of contribution is about 10\% of the total
torques on the first layer close to injection interface. (ii) Cancellation
effect among different $\mathbf{k}_{\parallel}$ in 2D BZ \cite{MDstiles02}.

\begin{figure}[tbp]
\includegraphics[width=8.6cm]{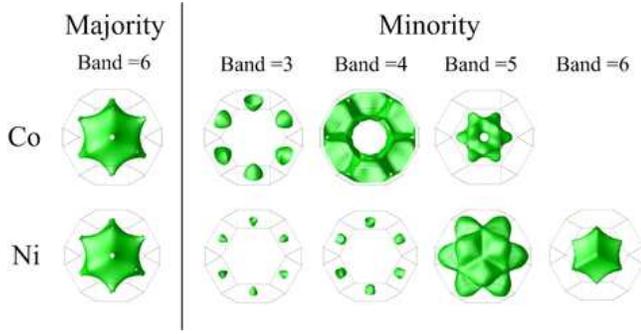}
\caption{(color online) The First row: Fermi surface projection of
the Co bulk fcc Brillouin zone on to a plane perpendicular the (111)
direction. Left-hand panel is for majority electron and right-hand
panel is for minority electron with band 3,4,5 FS. The Second row is
Ni bulk. } \label{fs}
\end{figure}

The materials dependency of the STT could be understood based on the Fermi
surface of Ni and Co. The wave vector $k_{\mu(\nu)}^{\downarrow(\uparrow)}$
can be found by the projection of minority spin (majority spin) Fermi
surface of ferromagnet along the current direction, where $\mu$ ($\nu$)
denotes the different sheets of Fermi surface for minority spin (majority
spin). In Fig.\ref{fs}, the Fermi surface for Co and Ni viewed along the
(111) direction for majority and minority spins is shown. As the shape of
Fermi surface for majority spin and minority spin in Co are greatly
different to each other, the precession frequencies $\bigtriangleup k_{\mu
\nu }$ of injected spin are varied rapidly as $\mathbf{k}_{\parallel}$
running over the 2DBZ. After summation of 2D BZ, the strong cancellation is
expected as shown in Fig.\ref{torque3}(a) and Fig.\ref{coni}(a). However,
for Cu$|$Ni, due to the similar symmetry between the wave function of the
sheet $\mu =6$ of minority spin and that of Cu,the electrons pass Cu$|$Ni
interface mainly through this channel($\mu =6$). While for majority spin,
the Fermi surface contains only one sheet. The precession frequency is
dominated by $\bigtriangleup k_{6 6 }$. Similar shape of sheet $\mu =6$ of
minority spin and sheet $\nu =6$ of majority spin will result in amount of
propagating states with similar precession frequency. After summation of
those states, the cancellation could be week and collective oscillation must
be of long period.

The above physical picture about spin torques in FM should be qualitatively
applicable to the system with FM to be Ni$_{80}$Fe$_{20}$. For Ni$_{80}$Fe$%
_{20}$, the overall band structure resembles that of Ni, however, due to the
scattering of Fe impurity atoms, the fine structure at Fermi surface could
be much more complicated than that of Ni. The dispersion of precession
frequency $\bigtriangleup k_{\mu \nu }$ is large and the cancellation should
be strong. As a result, the decay of spin torque is much faster than in the
conventional FM, Ni and Co. Besides, spin-orbit coupling is not included in
our calculation yet, which could introduce new mechanism of decay in Ni$%
_{80} $Fe$_{20}$.

\subsection{Interfacial disorder}

\begin{figure}[tbp]
\includegraphics[width=8cm]{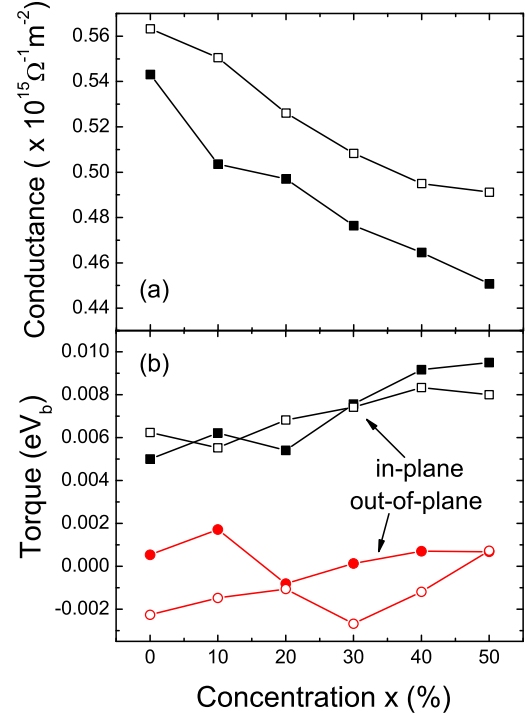}
\caption{(color online) The concentration $x$ dependence of the conductance
(a) and total spin torques on the free magnet (b) of spin valve Co($\protect%
\theta =90$)$|$Cu$|$FM$|$Cu with interfacial alloy
Cu$_{x}$FM$_{1-x}$, where the solid symbol for FM to be Co and open
symbol for FM to be Ni. In (b), the black symbol for in-plane
torques and red symbol for out-of-plane torque. }
\label{codisordercondx}
\end{figure}

\begin{figure}[tbp]
\includegraphics[width=9cm]{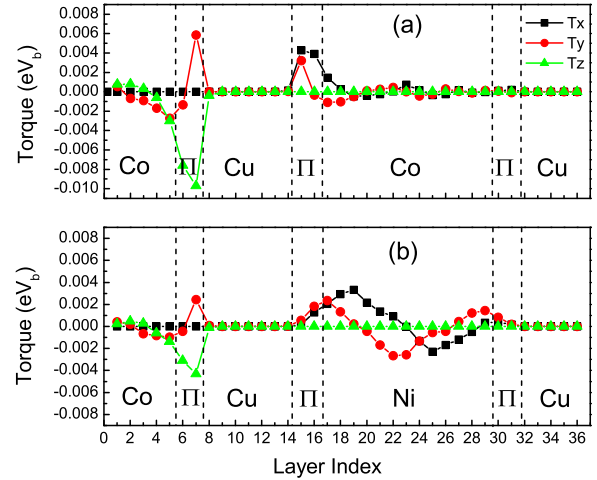}
\caption{(color online) The layer dependence of STT on interfacial
unit cell in the spin valve Co($\protect\theta =90$)$|$Cu$|$FM$|$Cu
with interfacial disorder, (a) FM is Co, (b) FM is Ni. The zone
labelled by $\Pi$ is the layers with substitutional alloy
Cu$_{50}$FM$_{50}$.} \label{codisorder}
\end{figure}

Interfacial disorder is likely to exist in the metallic system. Previous
studies\cite{Xia06} showed that the interfacial alloy could change the
polarization of the interface resistance. How the interfacial disorder
affects the STT is question we would like to answer in this section. The
interfacial disorder is introduced by two layers of substitutional alloy Cu$%
_{x}$FM$_{1-x}$ and Cu$_{1-x}$FM$_{x}$at interface between Cu and FM. In
present study, alloy is modelled by a $8\times 8$ lateral supercell, which
was shown to be a good modelling of the interfacial alloy\cite{Xia06}.

\begin{figure}[tbp]
\includegraphics[width=9cm]{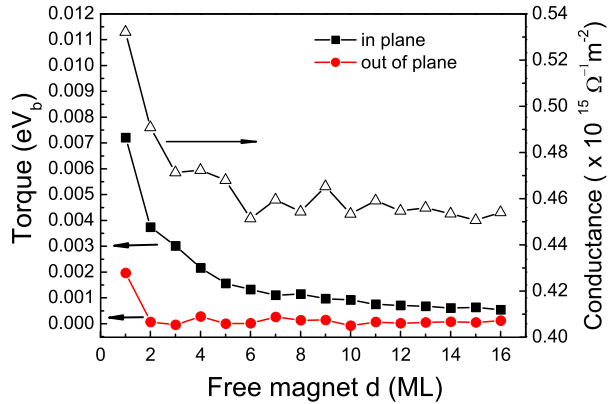}
\caption{(color online) The thickness $d$ dependence of the averaged
spin torques on the atom in free magnet and the total conductance of
the spin valve.} \label{cuco_disord_layer}
\end{figure}

In Fig.\ref{codisordercondx}(a)\&(b), the concentration $x$ dependence of
total conductance and total torque(in-plane and out-of-plane) on free magnet
are given for the realistic spin valve of Co($\theta=90^{o}$)$|$Cu$|$Co(Ni)$%
| $Cu. The total conductance decreases with increase of the concentration $x$%
, which means the interfacial disorder will suppress electronic transport in
this system. However, as shown in Fig.\ref{codisordercondx}(b), the total
in-plane torques acted on free magnet increase when the disorder is enhanced
in spite of the almost constant out-of-plane spin torque. The increase is
around 50\% for Co and 30\% for Ni.

In Fig.\ref{codisorder}, the layer dependence of spin torques is shown.
Comparing with Fig.\ref{torque3}(a), it is found that the torques on atoms
near the interface with Cu spacer have been enhanced. This result means that
the interfacial roughness will not kill the interfacial spin torques, on the
contrary, the dirty interface may be helpful to enhance the torques transfer.

For the spin valve with FM is Co, the free magnet thickness $d$ dependence
of the total conductance and STT are shown in Fig.\ref{cuco_disord_layer},
where the spin torques is obtained by average of total torques on free
magnet over all atoms in free magnet. Due to the quantum size effect, the
conductance decays with a small oscillation and tend to be constant with
increase of free magnet thickness. The in-plane torques dominates over the
out-of-plane torques for all thickness.

\section{Summary}

Based on the first principles frame, a method was developed to calculate the
transport and spin torques of the layered system with noncollinear
magnetization in linear response regime. STT in the ferromagnetic (FM) spin
valves are calculated. We found that the behavior of spin torques in the
free layer are greatly dependent on the materials. The cancellation effect
of the STT due to the different precessional frequency is sensitive to the
band structure of material. The contribution of evanescent states to the STT
is found to be nontrivial at the NM$|$FM interface. The effect due to
interfacial disorder is also considered, it is found that average torques
are enhanced in the presence of disorder.

\begin{acknowledgments}
This work is supported by NSF(10634070) and MOST(2006CB933000, 2006AA03Z402)
of China.We are grateful to: Paul Kelly for useful discussion; Ilja Turek
for his TB-LMTO-SGF layer code which we used to generate self-consistent
potentials; Anton Starikov for permission to use his version of the TB-MTO
code based upon sparse matrix techniques with which we can solve Eq.(\ref%
{coeff}) in an efficient way.
\end{acknowledgments}

\end{document}